%% file: HorvathCrisis.tex
\documentclass[12pt]{article}
\usepackage{epsfig}

\input econfmacros-csqcd3.tex
\def\Title#1{\begin{center} {\Large {\bf #1} } \end{center}}

\begin{document}

\Title{Is There a Crisis in Neutron Star Physics?}

\bigskip\bigskip


\begin{raggedright}
{\it J.E. Horvath\index{Horvath, J.E.}\\
Instituto de Astronomia, Geof\'isica e Ci\^encias Atmosf\'ericas\\
Universidade de S\~ao Paulo\\
05570-010 Cidade Universit\'aria\\
S\~ao Paulo, SP\\
Brazil\\
{\tt Email: foton@astro.iag.usp.br}}
\bigskip\bigskip

\noindent{and}

{\it O.G. Benvenuto\index{Benvenuto, O.G.}\\
Facultad de Ciencias Astron\'omicas y Geof\'isicas\\
Universidad Nacional de La Plata\\
Paseo del Bosque S/N\\
1900 La Plata, Bs. As.\\
Argentina\\
{\tt Email: obenvenuto@fcaglp.unlp.edu.ar}}
\bigskip\bigskip
\end{raggedright}

\section{Introduction}

Sometimes it becomes difficult for people not following the development
of a field to judge the advances and ongoing changes. Neutron star physics
is no exception to this rule. As a specific example of how long-lasting
concepts and results can be fundamentally changed, we may quote the
standard statement in all textbooks and even scientific articles up
to the beginning of the 21st century which reads: neutron star
masses are compatible
with a single scale of $1.4 M_{\odot}$. Mounting evidence for massive
and light neutron stars was considered as unreliable and unconfirmed.
And despite the possible width of this ``single-distribution'', the
belief that nature had some robust mechanism to produce quasi-identical
neutron stars was strong and is still present.

However, the situation changed dramatically in the last years,
when reliable new evidence become available. On the one hand, very
precise reports using different methods~\cite{Demorest,
Antoniadis} have demonstrated the existence of objects with $\sim
2 M_{\odot}$. On the other, low masses have been reported for a
few systems ~\cite{Meer}, even below the absolute minimum expected
theoretically for any neutron star formed in a supernova, although
systematic errors can not be completely controlled as yet.

We shall show in this report that the theoretical evolution of a
particular class of systems containing a neutron star,
the so-called ``black widow'' binaries, suggest masses above the
$\sim 2 M_{\odot}$, a fact that would bring serious concerns about
the right description of the equation of state above the
saturation density. Moreover, so far the actual determinations of masses
for these systems consistently give values above $2 M_{\odot}$,
reinforcing the quandary. In this sense, and given that the confirmation
of these ideas would create problems for the microphysical
description, we argue that a crisis
could be ``in the works'' in neutron star physics.

\section{Stellar Evolution and Masses}

It is now quite apparent that there are neutron stars with masses
above and below the ``canonical'' value of $1.4 M_{\odot}$. A few
recent works using the database by Lattimer, Steiner and
collaborators ~\cite{Jim} have concluded that at least two
different scales are present ~\cite{Turca, Zhang, Kiziltan,
Valentim}, although the exact values and physical origins remain
controversial. The suggestion by ven den Heuvel ~\cite{Ed} that
massive neutron stars may arise from massive cores developed by
stars with $M \geq 18 M_{\odot}$ (also present in the calculations
of Timmes, Woosley and Weaver ~\cite{TWW})could be related to a ``direct''
channel of formation. Naturally, accretion histories in LMXB and
related objects can also contribute to produce heavy objects.
The celebrated detection of a
$\sim 2 M_{\odot}$ by direct measurement of the Shapiro delay by
Demorest et al. ~\cite{Demorest} stands firm, but not unique,
among the high-end numbers (see below).

After the discovery of the first ``millisecond'' pulsar in 1982
~\cite{Backer}, another striking discovery came to enlarge the
variety of neutron star systems ~\cite{Frutcher} with the addition
of the first member of the ``black widow'' class. These systems
were rapidly identified as ablating the matter from the companions,
and therefore killing them much in the same way the homonymous
spiders do. In fact, low masses were measured for the latter, and
the presence of matter coming out of the system confirmed the
basic picture. The latest additions to the group include the
system PSR J1719-1438 in a 2.2 h orbit featuring a Jupiter-mass
companion ~\cite{Bailes} and PSR J1311-3430, discovered in
gamma-rays first by Pletsch et al. ~\cite{Pletsch} and
almost completely devoid of
hydrogen. However, it did not become clear how these systems are
formed, since their evolutionary paths could not be identified or
calculated with certainty. Black widows (and their cousins in which
the donors have higher masses, named after the
australian ``red backs'', see ~\cite{Roberts}) were loosely
identified as relatives of the LMXB, but their exact relationship
remained obscure. The follow-up of these ideas by actual numerical
calculations recently gave an unexpected bonus as we shall see
below.

\section{Calculations}

We have performed numerical calculations to understand the
conditions of formation of these black widow systems. A full
description of the Henyey code solving simultaneously the stellar
structure + stellar orbit equations is given in Benvenuto \& De
Vito (2003). The application to these specific systems has taken
into account 1) the detailed behaviour of the Roche Lobe region;
2) the evaporating wind $\propto L_{P} {(R_{2}/{a})}^{2}$ (with
$R_{2}$ the radius of the donor, $L_{P}$ the luminosity of the
pulsar and $a$ the semiaxis of the orbit, see ~\cite{Stevens})
although irradiation feedback was discarded because of simplicity.
We have
generated several tracks starting from a quite narrow period
interval (otherwise, the orbit could widen immediately and a black
widow will not be produced, see ~\cite{BDH}) and several choices
of the initial donor mass within accretion stability limits. By
hypothesis, we have chosen to start with a just-formed neutron
star of $1.4 M_{\odot}$ and assumed a fixed value of the transfer
efficiency ${\dot{M_{1}}} = - \beta {\dot{M_{2}}}$, although this
parameter is not really crucial for the outcome.

After $\sim few \, Gyr$ transfer the orbit shrinks when the donor star
becomes semi-degenerate and the wind from the pulsar ablates matter until
short period/low masses systems, as the ones observed follow. However,
the most relevant
result for our discussion here is the final NS mass. Fig.1 displays
the evolution of both the donor and the NS masses along the evolutionary
history of the system for a fixed value of $\beta = 1/2$. We see that
a good amount of accretion onto the NS is in fact very important for the
evolution of the systems, otherwise they could not occupy the region
of short period/low masses as observed. Therefore, and unless an actual
calculation with variable $\beta (t)$ prove otherwise, even a modest
amount of efficiency would drive the NS above the $2 M_{\odot}$ value.
This is quite striking but sound, since a long history of accretion
with episodes of ablation must have an effect on the accretor after all.

It was not then unexpected that very recent work combining
photometry and spectroscopy for the system PSR J1311-3430 ~\cite{Romani}
rendered high values for the NS mass, namely $2.15 \pm 0.11 M_{\odot}$,
$2.68 \pm 0.14 M_{\odot}$ or even $2.92 \pm 0.16 M_{\odot}$, depending
on the interpretation of the light curve. It is also important to remark
that the ``original'' black widow, PSR 1957+20 NS mass has been estimated
as $2.4 \pm 0.12 M_{\odot}$ before ~\cite{Kul}, in complete agreement with
the theoretical calculations. While there may be still a fine adjustment
of the final masses, we may have to consider seriously the existence of
masses even larger than the Demorest et al. (2012) determination.

\begin{figure}[htb]
\begin{center}
\epsfig{file=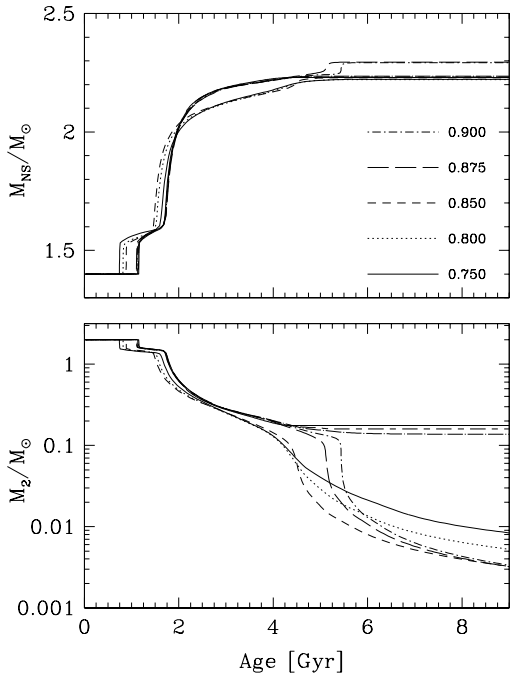,height=3.5 in} \caption{The mass of the
neutron star in the system PSR J1311-3430 (upper panel) and its
companion (lower panel). Several $Gyr$ are needed to place the
system in the observed orbit with the observed value of the donor
mass, and if the efficiency of accretion is not absurdly low, this
history will also produce a massive neutron star.}
\label{fig:mass}
\end{center}
\end{figure}

\section{Discussion}

Is there a crisis in NS physics after all? If the emerging values of
the black widow NS are confirmed, the answer will be ``yes''.
After many years standing on the $1.4 M_{\odot}$, the evidence for larger
masses seems overwhelming, and the work on the black widow systems
push the maximum value to ``uncomfortable'' levels. The answer will not
come easily, although the recent work on the hyperon sector suggest
there is unexpected repulsion hidden there, leading to consider a variety
of models in which hyperons could stand masses $> 2 M_{\odot}$ instead
of turning the stellar sequences downwards. The exotic models resorting to
quark matter, even in its extreme SQM version would also be insufficient
since in both popular versions (MIT bag and NJL, both with CFL-type pairing)
the mass cannot be increased indefinitely by tuning the parameters
~\cite{Paulucci}.

We must add here that there is a real possibility that rapidly rotating neutron stars
could increase their masses by $\sim 20 \%$ near Keplerian frequency. However,
the very fast rotation of both quoted black widow pulsars ($2.5 ms$ for
PSR J1311−3430 and $1.6 ms$ for PSR 1957+20) are still fast from this extreme
condition (see, for example, Weber, Orsaria and Negreiros, these Proceedings),
and their ``extra'' mass is not larger than a few percent for any type of
equation of state.

We shall witness a new round of theoretical work to
see how large masses are accommodated after all, although nature surely knows
how as shown by the plain observations.

\bigskip
JEH acknowledges the Fapesp and CNPq Agancies, Brazil for financial support.

\end{document}

%% file: econfmacros-csqcd3.tex
\def\beq{\begin{equation}}
\def\eeq#1{\label{#1}\end{equation}}
\def\eeqn{\end{equation}}

\def\beqa{\begin{eqnarray}}
\def\eeqa#1{\label{#1}\end{eqnarray}}
\def\eeqan{\end{eqnarray}}

\let\bar=\overbar

\def\Dslash{\not{\hbox{\kern-4pt $D$}}}
\def\dslash{\not{\hbox{\kern-2pt $\del$}}}

\def\msb{{\bar{\ssstyle M \kern -1pt S}}}

\usepackage{fancyhdr,graphicx}